\documentclass[sigconf]{acmart}
\usepackage{csquotes}

\AtBeginDocument{%
  \providecommand\BibTeX{{%
    \normalfont B\kern-0.5em{\scshape i\kern-0.25em b}\kern-0.8em\TeX}}}

\copyrightyear{2021} 
\acmYear{2021} 
\acmBooktitle{8th International Workshop on Middleware and Applications for the Internet of Things (M4IoT'21), December 6--10, 2021, Virtual Event, Canada}
\acmPrice{15.00}
\acmDOI{10.1145/3493369.3493601}
\acmISBN{978-1-4503-9167-2/21/12}

\acmConference[M4IoT'21]{8th International Workshop on Middleware and Applications for the Internet of Things}{December 6--10, 2021}{Virtual Event, Canada}
\setcopyright{none}
\renewcommand\footnotetextcopyrightpermission[1]{}
\usepackage{textpos}
\usepackage{framed}

\begin{document}


\title[RedCASTLE: Practically Applicable $k_s$-Anonymity for IoT Streaming Data at the Edge in Node-RED]{RedCASTLE: Practically Applicable $k_s$-Anonymity \\for IoT Streaming Data at the Edge in Node-RED} %

\author{Frank Pallas}
\email{fp@ise.tu-berlin.de}
\orcid{0000-0002-5543-0265}
\affiliation{%
  \institution{TU Berlin, Information Systems Engineering}
  \city{Berlin}
  \country{Germany}
}

\author{Julian Legler}
\email{julian.legler@campus.tu-berlin.de}
\orcid{0000-0001-8715-0156}
\affiliation{%
  \institution{TU Berlin}
  \city{Berlin}
  \country{Germany}
}

\author{Niklas Amslgruber}
\email{n.amslgruber@campus.tu-berlin.de}
\affiliation{%
  \institution{TU Berlin}
  \city{Berlin}
  \country{Germany}
}

\author{Elias Grünewald}
\email{eg@ise.tu-berlin.de}
\orcid{0000-0001-9076-9240}
\affiliation{%
  \institution{TU Berlin, Information Systems Engineering}
  \city{Berlin}
  \country{Germany}
}

\renewcommand{\shortauthors}{Pallas, Legler, Amslgruber, and Grünewald}

\begin{abstract}
  In this paper, we present RedCASTLE, a practically applicable solution for Edge-based $k_s$-anonymization of IoT streaming data in Node-RED. RedCASTLE builds upon a pre-existing, rudimentary implementation of the CASTLE algorithm and significantly extends it with functionalities indispensable for real-world IoT scenarios. In addition, RedCASTLE provides an abstraction layer for smoothly integrating $k_s$-anonymization into Node-RED, a visually programmable middleware for streaming dataflows widely used in Edge-based IoT scenarios. Last but not least, RedCASTLE also provides further capabilities for basic information reduction that complement $k_s$-anonymization in the privacy-friendly implementation of usecases involving IoT streaming data. A preliminary performance assessment finds that RedCASTLE comes with reasonable overheads and demonstrates its practical viability.
\end{abstract}


\maketitle

    \begin{textblock*}{\textwidth}(0cm,-18.5cm) 
    \begin{center}
    \begin{framed}
        \textit{preprint version (2021-10-29), accepted as Regular research paper for the\\ \textbf{8th International Workshop on Middleware and Applications for the Internet of Things}\\ \copyright~2021 ACM, to be published in the ACM Digital Library: \url{https://dl.acm.org/doi/10.1145/3493369.3493601}}
    \end{framed}
        
    \end{center}
    \end{textblock*}

\section{Introduction}

Ensuring privacy is one of the most important challenges when implementing real-world IoT scenarios such as building automation, connected cities, intelligent energy grids, or smart healthcare \cite{zhou2017iot}. In all these and many further cases, the data collected and processed may reveal personal and sometimes highly sensitive information in a multitude of ways. %
Regulatory provisions such as the GDPR as well as users' demands to protect their privacy %
therefore often require to minimize the level of detail at which continuously flowing IoT data such as sensor measurements or observed events are processed. At the same time, respective data must still be detailed enough to facilitate intended functionalities.%

For balancing these often diverging goals, %
advanced anony\-mization techniques ensuring properties such as $k$-anonymity \cite{sweeney_k-anonymity_2002}, $\ell$-diversity \cite{machanavajjhala_l-diversity_2007}, or $t$-closeness \cite{li_t-closeness_2007} %
have been established. Algorithms and respective implementations for ensuring these are publicly available. However, their practical adoption in real-world IoT scenarios is currently hindered by at least two shortcomings: 

First, most algorithms and implementations focus on ano\-ny\-mi\-zing %
persistent datasets stored in and retrieved from, for instance, a database while IoT scenarios strongly rest upon streaming data that are processed ``on the fly''. Ensuring above-mentioned privacy properties for data streams, in turn, has seen significantly less coverage in the scientific discourse so far \cite{castle-paper,robinson_castleguard_2020}. Second, and similarly important, existing implementations are rather rudimentary and %
hardly address questions of integration into established tools and solutions employed in practice. This is particularly true for combining anonymization measures with Edge-computing approaches, which have also been proposed for privacy-friendly designs of IoT scenarios \cite{pallas-bermbach-raschke-fog}. Without such integration, however, respective anonymization techniques will hardly be adopted in practice.

To address these challenges, we herein propose RedCASTLE, an easily integratable and flexibly configurable anonymization extension to the widely used IoT middleware Node-RED. RedCASTLE allows to implement a broad variety of information reduction approaches as well as to ensure $k_s$-anonymity %
via the established CASTLE algorithm \cite{castle-paper} for IoT data streams with minimum integration effort. %
In particular, RedCASTLE comprises:

\begin{itemize}
    \item a set of configurable functions for basic information reduction, such as attribute suppression, data filtering, and data mapping,
    \item an extension of the pre-existing CASTLEGUARD library to facilitate the $k$-anonymization of actual data streams with numerical and non-numerical data, and
    \item a practically applicable and easily adoptable extension that coherently integrates said functionality into Node-RED, a highly interoperable middleware for streaming dataflows widely established for IoT- and Edge-usecases.
\end{itemize}

The provided extension is publicly available under an open source license.\footnote{RedCASTLE is available under the MIT license at \url{https://github.com/PrivacyEngineering/RedCASTLE}.} Our work builds upon a pre-existing, rudimentary implementation, CASTLEGUARD \cite{robinson_castleguard_2020}. %
So far, however, CASTLEGUARD is significantly limited in matters of practical applicability and lacks, for instance, connectivity to real streaming data sources (instead of merely simulating them by reading a .csv-file line by line), capabilities to handle non-numerical data, or integration into real-world pipelines. %
We herein address these limitations and thereby provide -- to the best of our knowledge -- the first 
solution for Edge-based $k$-anonymization of streaming data that is practically applicable in real-world settings.%

Our considerations and contributions unfold as follows: In section \ref{bg_rel}, we provide relevant background knowledge and related work. %
On this basis, we depict our general integration approach (section \ref{integration}) as well as our newly introduced functionalities for basic information reduction (section \ref{generalization}) and $k_s$-anonymization (section \ref{node-anon}) in Node-RED. A preliminary performance assessment of our solution is provided in section \ref{performance}, section \ref{conclusion} concludes.

\section{Background \& Related Work}\label{bg_rel}

In this section, we begin with the relevant background on IoT and the anonymization of streaming data in the Node-RED middleware.

\subsection{IoT, Edge, and the Role of Streaming Data}

Internet-of-Things (IoT) scenarios arise in a broad variety of applications, from building automation \cite{lilis2017}, connected cities \cite{liu2019intelligent}, or energy grids \cite{Samie2019} to smart healthcare \cite{pace-edge-health}. All these scenarios rest upon vast amounts of status and measurement data being collected, processed, integrated, and acted upon in a timely manner. 

After initial trends towards centralized, often cloud-based architectures where data are sent back and forth between the place of collection and effectuation (e.g., a smart meter collecting consumption data and controlling the charging behavior of an electric vehicle) on the one and a centralized processing pipeline on the other hand, recent developments increasingly recognize the need for more decentralized architectures. This is particularly driven by growing amounts of data conflicting with limited bandwidths and processing capacities and by near-real-time requirements of certain IoT usecases conflicting with inevitable network latencies of cloud-centric approaches. 

In Edge and Fog computing \cite{bermbach-fog} models, parts of the data processing are therefore carried out closer to the points of collection and effectuation. This allows to filter, aggregate, and otherwise preprocess data before forwarding them to upstream services and to implement significant parts of the functionality locally. %
Especially for continuous streams of measurement data and events from large numbers of sensors and devices, this significantly decreases the amount of data to be transferred as well as round-trip latencies between an event occurring and the respective response being carried out. Besides such possible benefits in matters of performance, Edge and Fog computing may, last but not least, also serve as enabling technology for more privacy-friendly implementations of IoT-scenarios through patterns such as early filtering, aggregation, or anonymization \cite{pallas-bermbach-raschke-fog}.

\subsection{Anonymization for Streaming Data}\label{stream-anon}

Data anonymization is one of the most fundamental techniques for implementing privacy-friendly systems. Na\"{i}ve approaches for doing so, %
however, pose the risk of re-identifiablity of individuals through so-called quasi-identifiers \cite{sweeney_k-anonymity_2002} and, thus, the factual disclosure of personal data. %
To avoid such risks, %
advanced anonymization schemes and measures like $k$-anonymity \cite{sweeney_k-anonymity_2002}, $\ell$-diversity \cite{machanavajjhala_l-diversity_2007} or $t$-closeness \cite{li_t-closeness_2007} have been established. %

Respective approaches are, though, designed with rather static datasets in mind and do not fit the givens and requirements arising in the context of IoT streaming data \cite{otgo18}. Besides the relatively slow anonymization process, which conflicts with near-real-time requirements in IoT usecases \cite{castle-paper}, this is particularly the case because the underlying assumptions do not hold for IoT streaming data. Instead, appropriately adapted anonymization models are required. 

Focusing on %
$k$-ano\-nym\-ity, a suitable model is model is $k_s$-ano\-nym\-ity, as implemented in the CASTLE algorithm \cite{castle-paper}. Here, arriving streaming data is assigned to different clusters based on automatically generalized values for a manually defined set of numerical quasi-identifiers. For instance, the algorithm may determine four different value-ranges for a quasi-identifier ``vendor-id'' and six value-ranges for a quasi-identifier ``station-id'' in an electric vehicle charging use case. All messages with similar combinations of so-generalized quasi-identifiers are then combined into one cluster.\footnote{Details on quasi-identifiers, their importance in the context of anonymization, etc. had to be left out due to space constraints. For more in-depth elaborations, see \cite{sweeney_k-anonymity_2002}.} 

Every cluster is then considered to be $k_s$-anonymous if it contains at least $k$ values. If clusters cannot be made $k_s$-anonymous, they will be merged with other clusters. When new data arrives and does not fit into an existing cluster, the closest cluster gets enlarged (but only if it is not already $k_s$-anonymous) or a new cluster is created.

Compared to other adopted models such as FAANST  \cite{zake11} or K-VARP \cite{otgo18}, CASTLE has seen the strongest recognition in the scientific discourse. In addition, a %
continuously maintained reference implementation is available as part of the CASTLEGUARD library \cite{robinson_castleguard_2020}. We therefore chose CASTLE as the basis for our streaming data anonymization component.

\subsection{Node-RED}

Node-RED is best described as a visually programmable middleware for streaming dataflows. It supports a broad variety of interfaces for data in- and e-gress and is widely used in several industries  for implementing complex and dynamically adaptable IoT data flows.\footnote{Existing large-scale industrial applications of NodeRED mentioned in developer forums include, for instance, medical settings, energy provision, or integration of industrial PLCs. See \url{https://discourse.nodered.org/t/node-red-at-enterprise-level/11205/8}} Given its low footprint, Node-RED is particularly suitable and advocated for usecases involving above-mentioned Edge-based preprocessing of IoT streaming data. %

Within Node-RED, all functionalities are provided via so-called \emph{nodes} which are dynamically linked through \emph{wires} into \emph{flows}. Messages are brought into a flow via special \emph{input} nodes, which exist for a broad variety of data sources such as MQTT channels,  %
message buses or mesh networks like KNX or ZWave, or even low-level UDP datagrams. Similarly, %
messages can be published via \emph{output} nodes, which can again represent an MQTT channel, an HTTP call to be performed, etc. Between input and output, messages are processed and may be transformed in \emph{function} nodes of different kinds. In all three categories, available node types are manifold and the library is continuously extended.\footnote{For the all available node types, see \url{https://flows.nodered.org/search?type=node}.} %

The broad spectrum of available node types notwithstanding, anonymization capabilities -- especially following advanced schemes like $k_s$-anonymity -- are currently lacking in the Node-RED ecosystem. This particularly hinders the adoption of Node-RED in privacy-sensitive usecases or requires separate, non-integrated measures like anonymization proxies to be added between a Node-RED instance and any upstream service. %
Both options would come with significant downsides in matters of implementable usecases, increased efforts, or performance drops.

Instead, we thus propose to integrate advanced anonymization capabilities directly into Node-RED. %

\section{Integrating Anonymization into Node-RED}\label{integration}

In line with other endeavors for practically applicable privacy engineering \cite{pallas-ea-2020-al-pbac,gruenewald2021, gruenewald2021tira}, our solution shall not only provide the functionality to $k_s$-anonymize data but also fulfill further nonfunctional requirements such as coherently integrating with established toolchains and respective application patterns or raising low integration effort. %

In this vein, RedCASTLE is provided as a self-contained extension to Node-RED that encapsulates the underlying functionality of CASTLEGUARD and makes it available through a custom function node that natively integrates and can be used in flows like any other function node. Similarly, %
functionalities for basic information reduction %
are also made accessible in a separate class of function nodes. Thereby, RedCASTLE decouples the anonymization functionality as far as possible from Node-RED's core and ensures future-proofness. As CASTLEGUARD is implemented in Python while Node-RED requires custom nodes to be written in JavaScript, messages are exchanged between these subcomponents via a low-overhead, brokerless local ZeroMQ message queue.

\begin{figure}[t]
    \centering
    \includegraphics[width=1.0\linewidth]{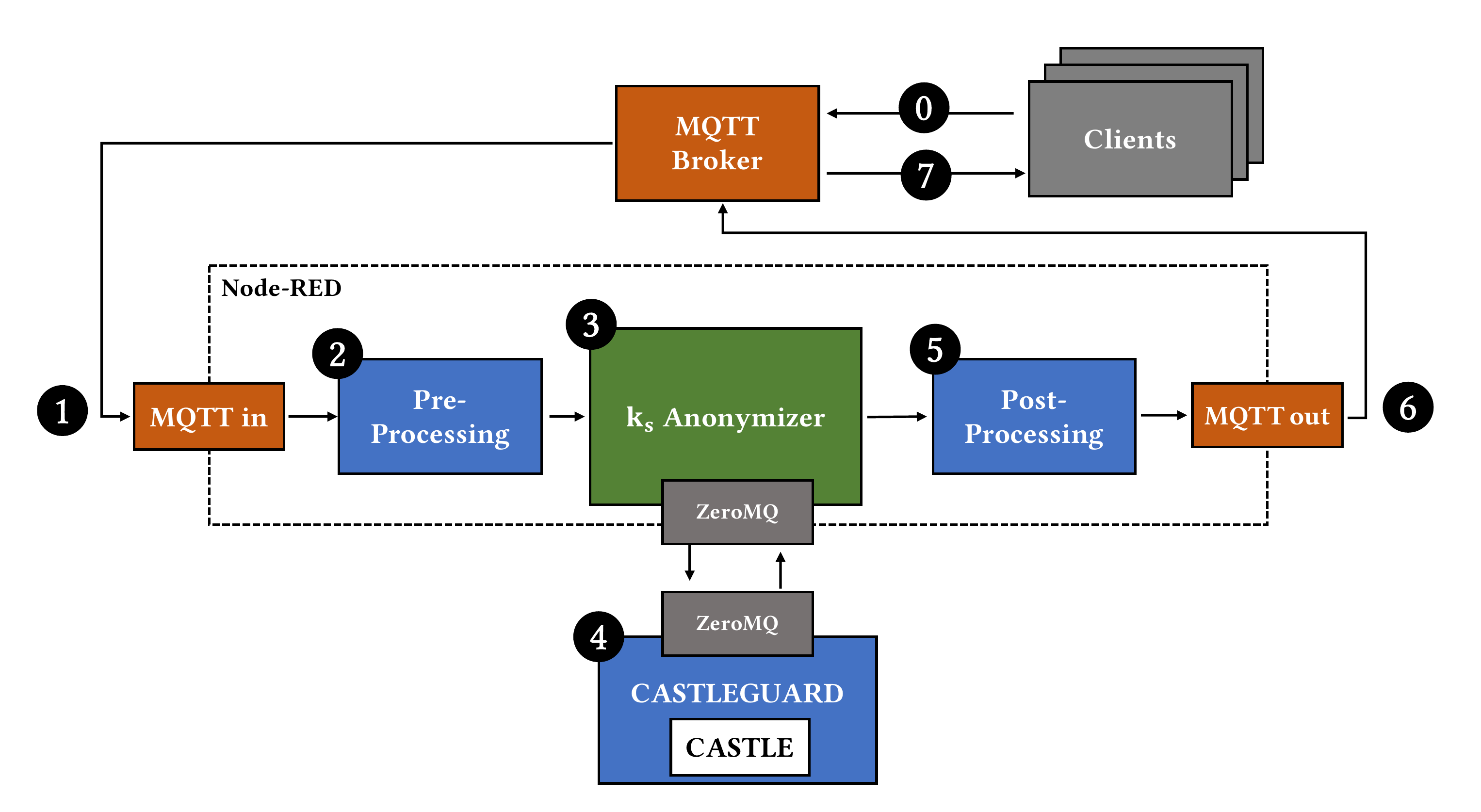} %
    \caption{$k_s$-anonymization process in RedCASTLE.}
    \label{fig:architecture}
\end{figure}

Based on these building blocks, $k_s$-anonymization of IoT streaming data can be implemented within Node-RED in line with its common visual programming paradigm and respectively established patterns and practices as follows (see figure \ref{fig:architecture}).

\paragraph{Ingress} The messages to be anonymized enter a flow through any kind of input node available in Node-RED, ensuring maximum flexibility and interoperability. A quite common usecase might here be an MQTT-input that subscribes to one or multiple channels.\vfill\null

\paragraph{Information Reduction} Even though not necessarily required for $k_s$-anonymization, performing basic information reductions -- attribute suppression, filtering, mappings -- on the messages beforehand allows to eliminate unnecessary but possibly privacy-sensitive attributes, reduces complexity for the subsequent step, and also helps rendering initially unfitting data suitable for automated generalization (e.g., when mapping a discrete vehicle model string to a numerical price parameter in a smart charging scenario). In addition, it may help reduce the bandwidth required for forwarding messages from an Edge-node to upstream services afterwards. Such reduction is done in an information reduction node that is added to the flow and wired to the input node. Available reductions as well as configuration parameters are laid out %
in section \ref{generalization} below. 

\paragraph{$k_s$-Anonymization} The actual generalization and clustering of messages according to the $k_s$-anonymization model laid out above (see section \ref{stream-anon}) is done with a separate CASTLEGUARD node. This node abstracts away the complex functionality of the underlying component (as well as respective inter-component communication) and is simply wired to the information reduction node and, thus, fed with pre-processed messages. %
Again, available functionalities and respective configuration parameters are laid out in more detail below in section \ref{node-anon}. As soon as a cluster fulfills the $k_s$-criterion, respective messages are bulk-released by the CASTLEGUARD node.

\paragraph{Re-Publishing} To forward anonymized data to upstream services outside of Node-RED like a cloud-based processing pipeline or subsequent Edge-local components, %
a respective output node is wired to the CASTLEGUARD node. Again, a quite common usecase might here be an MQTT-output publishing respective messages via an external broker. For doing so, the CASTLEGUARD node only needs to be wired to any of the output nodes available in Node-RED.

Of course, additional function nodes can be inserted at any stage of this basic flow: Before generalization takes place, before the actual $k_s$-anonymization, or after the CASTLEGUARD-step. Similarly, some use cases might only use an information reduction node and go without $k_s$-anonymization or vice versa. This way, our separated nodes integrate well into larger, more complex flows, allowing to flexibly work with anonymized data in Node-RED.

\section{Basic Information Reduction}\label{generalization}

For basic data reduction, RedCASTLE provides the following message manipulations that can all be configured by attaching the configuration to the specific message\footnote{For the specific syntax to be used for this and all subsequently described configurations, see \url{https://github.com/PrivacyEngineering/RedCASTLE}.}:

\paragraph{Attribute Suppression} Not all parameters of incoming messages may be relevant for the dataflow to be carried out. At the same time, removing certain attributes (such as individual identifiers) from messages may provide benefits in matters of privacy and/or required bandwidth. RedCASTLE therefore allows to specify names for those attributes that are to be stripped from every message that passes a data reduction node. For this aim, a \texttt{suppress properties} node allows to remove %
attributes from messages accordingly.%

\paragraph{Filters} Besides suppressing single attributes, messages can also be filtered out completely based on different conditions. In particular, RedCASTLE implements attribute-driven allow- and disallow filters. This does, for instance, completely drop all messages with an \texttt{objectID}-value included in a provided set of disallowed IDs. In addition, with a range-filter numerical value ranges for message attributes can be specified. In this case, all messages with the respective values being outside the respective range are dropped.

\paragraph{Conditional Changes} Conditional changes basically allow to manipulate message data depending on conditions being matched or not. RedCASTLE allows to add or change %
the value of an attribute \texttt{changeAttributeName}, either on the basis of a string being matched or based on numerical value ranges. This allows, for instance, to implement above-mentioned mapping functionality: whenever a parameter \texttt{vehicle-model} matches a particular string, a new numerical parameter \texttt{vehicle-price} may be set. Similarly, numerical price ranges may also be explicitly mapped to (numbered or string-named) price-categories. 

Except range-based ones, all these reduction functions can be used with numerical and non-numerical attributes. Additional functionalities might be added in the future, but with suppression, filtering, and conditional changes, RedCASTLE already provides the most relevant capabilities for information reduction on continuously flowing messages.

\section{Node-RED-Adopted $k_s$-Anonymization}\label{node-anon}

Providing practically valuable $k_s$-anonymization in Node-RED based in the pre-existing CASTLEGUARD implementation required several extensions to be made. In particular, this regards the previous lack of suitable integration interfaces as well as missing support for non-numerical data.

\subsection{Integration Interface} First and foremost, we added actual streaming data interfaces as in- and outputs to CASTLEGUARD. Before, CASTLEGUARD only accepted .csv-files as inputs and printed $k_s$-anonymized outputs to the command-line, severely limiting its practical use for real-world scenarios. We therefore extended CASTLEGUARD by a lightweight ZeroMQ interface allowing for a coherent and low-overhead integration and message exchange with Node-RED. 

This interface is employed by our above-mentioned abstracting RedCASTLE function node, which basically receives a message within the Node-RED context, ensures that the modified CASTLEGUARD process is running, and forwards the message \enquote{as-is} to this process via said message queue. %
Similarly, whenever messages are bulk-released by CASTLEGUARD, this is also done via a second ZeroMQ interface listened to %
by the RedCASTLE function node. All respective, $k_s$-anonymized messages are then released by the function node and forwarded and processed within Node-RED as usual. Once RedCASTLE is installed, all this works seamlessly and automatically, without requiring any further configuration etc.

\subsection{Non-Numerical Data}\label{non-numerical} In addition, we also extended CASTELGUARD itself to provide advanced functionality for handling non-numerical data by automatically converting them to numerical categories. Non-numerical data could so far not be handled at all by the pre-existing implementation. Given that in real-world IoT scenarios, message attributes are non-numerical (e.g., string-based) quite often and that these attributes (such as, for instance, a vehicle model) might be relevant quasi-identifiers, this significantly limits practical applicability. %
To at least partially close this gap, we extended CASTLEGUARD with basic capabilities for handling non-numerical message attributes and for incorporating them in the $k_s$-anonymization process, including automated categorization.

For this purpose, \texttt{non-categorized-attributes} can be specified in RedCASTLE's configuration. Whenever a so far unseen value is detected for one of these attributes, it is assigned to a new numerical category ID and %
replaced in the message accordingly. A previously seen value, in turn, is replaced with the previously determined category so that, e.g., all occurences of a \texttt{vehicle-model} ``e-tron 55'' are replaced with the same category ID.

In CASTLEGUARD's $k_s$-anonymization procedure, these categories are then treated specifically. As there is no natural ordering of category IDs or, respectively, no semantic meaning embodied in their ordering, grouping them based on value ranges would not have made sense. Instead, categories are treated as sets in our extended implementation. Clusters may then be created independently from the category ID ordering and without generalizing them so that, for instance, one cluster may comprise the IDs $\{3,6\}$ and the other one $\{1,2,4,9\}$. Consequently, instead of min-max-ranges, a list of all category IDs inside a cluster is also placed into the output.

\subsection{Anonymization Parameters}\label{anon-params}%

Besides above-mentioned extensions and adaptations, we also made the underlying $k_s$-anonymization procedure highly configurable. Parameters are set in a JSON configuration file and can be divided into algorithm- and dataset-related ones.

\paragraph{Algorithm-specific parameters} In this group, parameters that control the $k_s$-anonymization procedure can be configured. This includes the \texttt{k} for the $k_s$-anonymity, the maximum amount of tuples \texttt{delta}, the maximum allowed active clusters \texttt{beta} and the configuration parameter \texttt{mu} for controlling the maximum information loss.

\paragraph{Dataset-specific parameters} For the algorithm to work correctly, some information has to be specified in this parameters group. The sensitive attribute has to be set as well as the quasi identifiers and, if existing, the identifier attribute. The attributes to be interpreted as non-numerical values as described in \ref{non-numerical} are also specified here.

\section{Preliminary Performance Assessment}\label{performance}

For validating at least the basic viability of our approach in matters of expectable overheads and to preclude being on a fundamentally flawed path, we conducted a set of preliminary performance assessments. 

In line with established best-practices for security- and privacy-related performance benchmarks \cite{Pallas2018Disillusion,pallas-ea-evidence-based}, we deployed 3 medium-sized n2-standard-2 Google Cloud instances %
to separate different components from each other. %
The first instance is used for the MQTT broker, the second is running the data emulator (the ``benchmarking client'') and the last one is running the actual system under test, the Node-RED server with RedCASTLE. To minimize external impact, all servers are created in the same availability zone and placed within a Virtual Private Cloud Network.

\begin{figure}[t]
    \centering
    \includegraphics[width=1\linewidth]{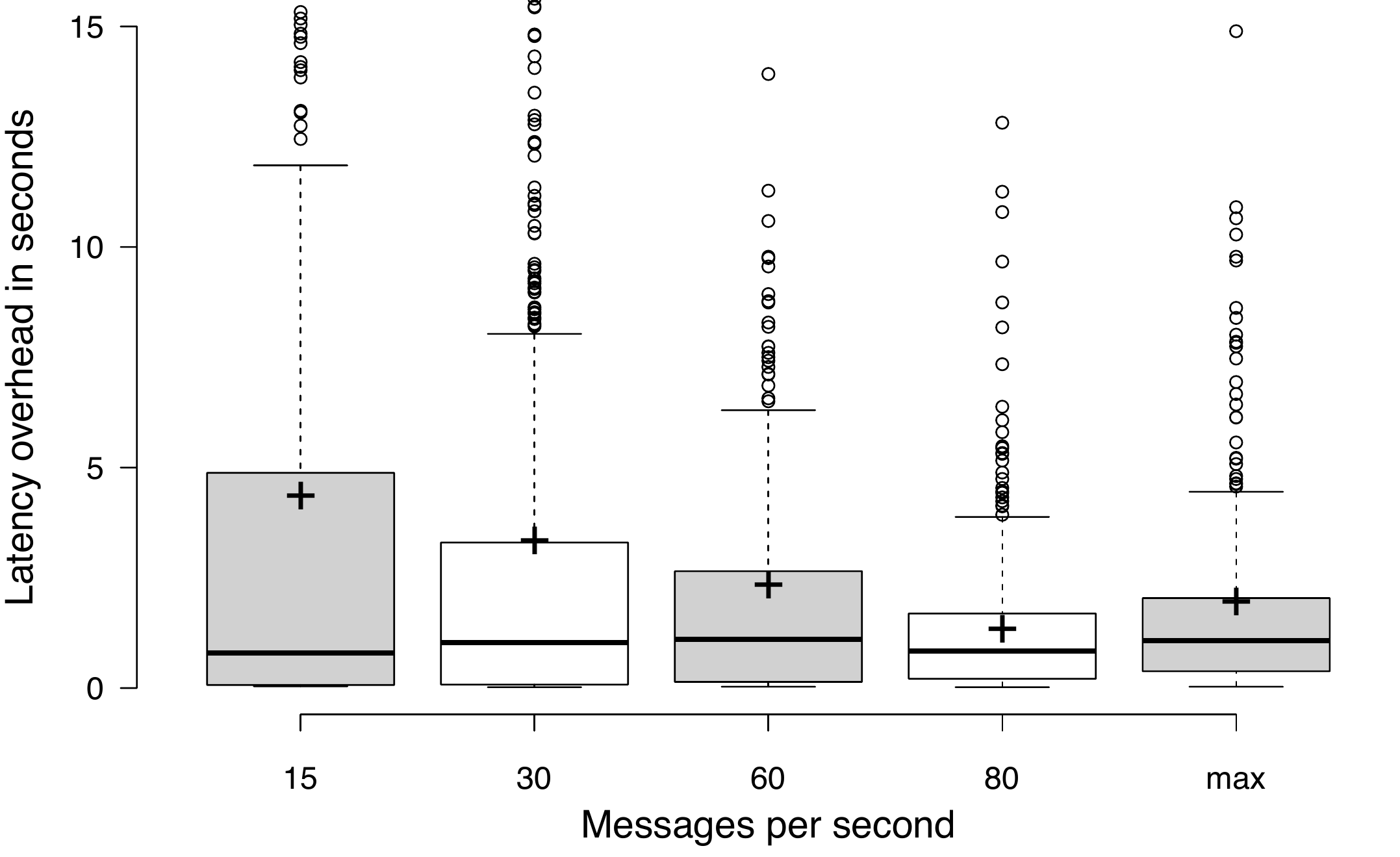} 
    \caption{Latency overhead induced by RedCASTLE's $k_s$-anonymization for different message frequencies (median, 25th and 75th percentile, and 1.5 times the interquartile range, outliers represented by dots, means by crosses).}
    \label{fig:overhead}
\end{figure}

Based on this general setting, we benchmarked 1) the additional delay introduced by RedCASTLE's $k_s$-anonymization %
and 2) the difference in matters of achievable message throughput with and without RedCASTLE being used. Benchmarks were conducted using a realistic dataset of electric vehicle charging events provided by the city of Boulder, Colorado.\footnote{"Electric Vehicle Charging Station Energy Consumption", \url{https://open-data.bouldercolorado.gov/datasets/183adc24880b41c4be9fd6a14eb6165f\_0/explore}%
}  %
To spice up the dataset, several fake persons with specific vehicle models and unique IDs were %
used to enrich the original dataset. We chose a realistic $k=5$ %
for our initial assessments, all other anonymization parameters (see \ref{anon-params}) were kept at their default.
All CPU- and network loads %
were constantly monitored during benchmark runs and stayed -- with one exception, see below -- within ranges ensuring we actually benchmarked what we intended to. %

\paragraph{Message delay} Given the clustering approach behind $k_s$-ano\-nym\-ity laid out above, messages are not immediately propagated through the message-flow defined in Node-RED but rather collected in a cluster until enough messages are present for successfully $k_s$-anonymizing them. This necessarily implies a delay of message delivery which can be expected to be higher for lower message frequencies.  We therefore determined the additional delays induced for 15, 30, 60, 80, and the highest possible amount of  messages per second. In line with our expectations, the medium delay per individual message as well as the observed deviations from this value decrease significantly with higher message frequencies, with the mean delay stabilizing around 1-2 seconds in our chosen scenario (see Fig. \ref{fig:overhead}). For many real-world usecases employing privacy-sensitive IoT streaming data, these results appear to be reasonable and acceptable. From 80 messages/s onward, however, the CPU load of one core increased significantly and also resulted in slight latency increases. Given Python's single-thread characteristics, this perfectly resembles RedCASTLE's expectable behavior and vividly illustrates the computational complexity behind $k_s$-anonymization.

\begin{figure}[t]
    \centering
    \includegraphics[width=1\linewidth]{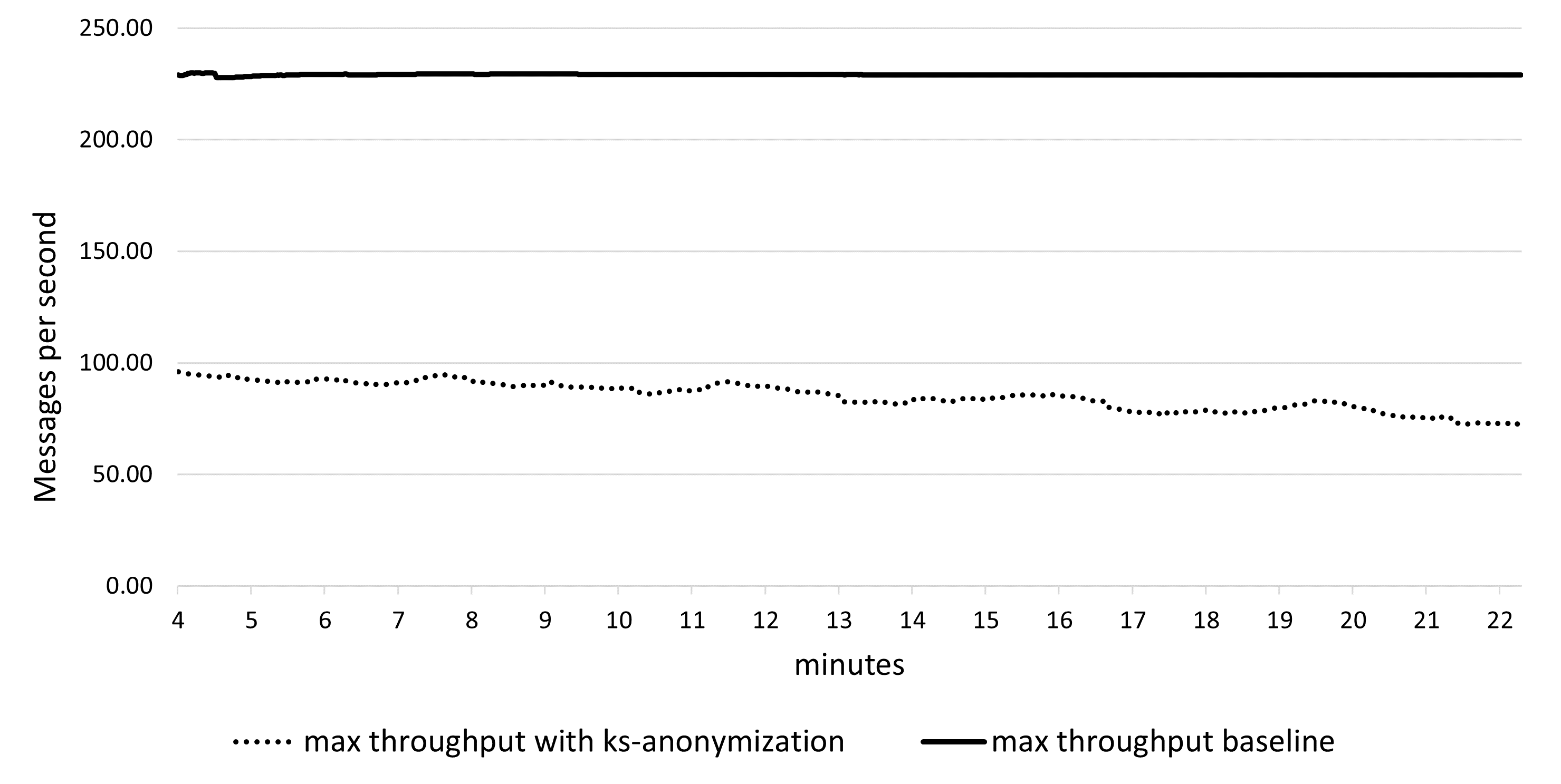} 
    \caption{Measured maximum throughput with and without $k_s$-anonymization component (moving 1-minute window)}
    \label{fig:throughput}
\end{figure}

\paragraph{Maximum throughput} Besides the delay necessarily introduced, our $k_s$-anonymization mechanism expectably also has an impact on the achievable message throughput, which we determined by letting %
our benchmarking client fire as many messages as possible. %
This resulted in a relatively stable average throughput around 90 messages/s with RedCASTLE's $k_s$-anonymization integrated into the Node-RED flow (see figure \ref{fig:throughput}). With $k_s$-anonymization being skipped, %
we were surprisingly no longer able to saturate the Node-RED instance: Message throughput in this case reached around 230 messages/s with neither the Node-RED instance nor the two other ones reaching a CPU load above 20\%. Network interfaces were also far from operating at full capacity. This points towards a so far unidentified bottleneck in Node-RED. %

On the one hand, this clearly indicates a strong need for further investigations to identify the actual bottleneck limiting message throughput. %
On the other, more pragmatically speaking one, the observed limitations are what Node-RED users are currently left with in the employed scenario, no matter what the bottleneck actually is. From this perspective, RedCASTLE's $k_s$-anonymization reduces achievable message throughput by roughly 60\% -- a significant overhead that will nonetheless be deemed reasonable in many real-world usecases and will also relativize with more complex and computationally intensive message flows being implemented at the Edge around RedCASTLE's anonymization.

Both the induced latencies and the throughput reduction do, finally, appear in a different light when taking into account that the anonymization functionality provided by RedCASTLE is indispensable for the lawful implementation of many real-world usecases involving privacy-sensitive IoT streaming data. When seen as such an enabling technology, the additional benefits that can be generated from respective usecase implementations will in most cases clearly outweigh or justify the observed overheads. Altogether, our initial performance assessments thus suggest non-negligible but still bearable overheads to result from applying RedCASTLE in real-world usecases. More in-depth investigations -- covering different values for $k$, more complex Node-RED flows %
and %
trying to pinpoint the observed bottleneck -- are nonetheless necessary for getting a more comprehensive picture in the future. By and large, however, RedCASTLE appears to be a practically viable approach for implementing indispensable anonymization functionality in Edge-based streaming data processing.

\section{Conclusion and Future Work}\label{conclusion}

In this paper, we presented RedCASTLE -- the, to the best of our knowledge, first practically viable solution for $k_s$-anonymization of IoT streaming data within a widespread Edge middleware. RedCASTLE builds upon a pre-existing, rudimentary implementation of the CASTLE algorithm, significantly extends it by actual streaming interfaces and capabilities for handling non-numerical data, and provides coherent integration into Node-RED. Due to the nature of $k_s$-anonymization and its computational complexity, RedCASTLE inevitably introduces overheads in matters of message latency and throughput. Given that many application scenarios will not be realizable without solid anonymization at all and that the observed overheads will relativize in more complex Edge data flows, %
these overheads will, however, be deemed bearable in many real-world usecases.

Interesting areas for future work include the addition of further information reduction functions as well as also transferring anonymization functionalities for $\ell$-diversity, $t$-closeness, and $\varepsilon$-differential privacy already provided by CASTLEGUARD. Implementing these in JavaScript instead of employing a pre-existing Python implementation, in turn, would expectably lead to a more native integration into Node-RED and eliminate kludges such as context-changing inter-process communication. Finally, more in-depth performance benchmarks as indicated in section \ref{performance} are also an advisable subject for future reasearch to better understand and optimize the impact of RedCASTLE in real-world scenarios.

These and further open issues notwithstanding, however, RedCASTLE is the first of its kind implementation for $k_s$-anonymizing IoT streaming data within a widely used Edge middleware in an integrated, coherent manner. It therefore allows for the lawful implementation of many respective usecases involving personal data and, thus, renders them practically viable at all.\vfill\null

\bibliographystyle{ACM-Reference-Format}
\bibliography{redcastle}

\end{document}